# A MULTIDISCIPLINARY APPROACH TO TELEGRAM DATA ANALYSIS

**Velizar Varbanov**[1]

**Kalin Kopanov**[1]

**Prof. Dr. Tatiana Atanasova**[1]

[1] Institute of Information and Communication Technologies - Bulgarian Academy of Sciences, **Bulgaria**

## ABSTRACT

This paper presents a multidisciplinary approach to analyzing data from Telegram for early warning information regarding cyber threats. With the proliferation of hacktivist groups utilizing Telegram to disseminate information regarding future cyberattacks or to boast about successful ones, the need for effective data analysis methods is paramount. The primary challenge lies in the vast number of channels and the overwhelming volume of data, necessitating advanced techniques for discerning pertinent risks amidst the noise. To address this challenge, we employ a combination of neural network architectures and traditional machine learning algorithms. These methods are utilized to classify and identify potential cyber threats within the Telegram data. Additionally, sentiment analysis and entity recognition techniques are incorporated to provide deeper insights into the nature and context of the communicated information.

The study evaluates the effectiveness of each method in detecting and categorizing cyber threats, comparing their performance and identifying areas for improvement. By leveraging these diverse analytical tools, we aim to enhance early warning systems for cyber threats, enabling more proactive responses to potential security breaches. This research contributes to the ongoing efforts to bolster cybersecurity measures in an increasingly interconnected digital landscape.

## INTRODUCTION

In the wake of recent geopolitical conflicts such as the Ukraine-Russia conflict and the Israel-Palestine tensions, the digital landscape has become increasingly fraught with cybersecurity risks [1, 2]. Hacktivist groups have leveraged platforms like Telegram to disseminate information regarding their intentions for cyberattacks or to celebrate successful exploits. However, with the vast data space in Telegram channels, distinguishing relevant cyber threat information from other data is a huge challenge.

To address this challenge, this study adopts a multidisciplinary approach, harnessing the power of advanced data analysis techniques to extract actionable insights from Telegram data. Our investigation focuses on channels associated with hacktivist groups. Notably, we observed two prominent topics dominating these channels: discussions pertaining to political events and discourse surrounding cyber threats [3].

Motivated by the need for effective data triage, we utilized two distinct datasets, each comprising approximately 4000 statements. These datasets were meticulously curated to facilitate the training of machine learning algorithms [3] and neural network architectures. By segregating the data according to thematic relevance – cyber threats and political discourse – we aimed to provide cybersecurity analysts with focused datasets tailored to their expertise.

The study evaluates the efficacy of three primary analytical methodologies: Feedforward Neural Network (FNN), Long Short-Term Memory (LSTM), and Support Vector Machine (SVM). Through rigorous testing and comparative analysis, we seek to elucidate the strengths and weaknesses of each approach in identifying and categorizing cyber threats within the Telegram data streams. Furthermore, to augment the discernment of valuable intelligence, we integrate entity recognition and sentiment analysis techniques [4, 5] into our analytical framework. By discerning key entities and gauging sentiment, we aim to provide nuanced contextual understanding, enabling more informed decision-making by cybersecurity experts and policymakers.

This research endeavour represents a concerted effort to fortify early warning systems against cyber threats in an increasingly volatile digital environment [6]. By leveraging interdisciplinary methodologies and innovative analytical techniques, we endeavour to empower stakeholders with actionable insights, thereby enhancing proactive cybersecurity measures and safeguarding digital infrastructures.

## MATERIALS AND METHODS

We progressed through various phases during the execution of our experiments:

**Data Collection** - It was gathered data from Telegram channels affiliated with hacktivist groups supporting various geopolitical causes, including Ukraine, Russia, Palestine, and Israel. This diverse selection of groups provided a comprehensive dataset reflecting a spectrum of political ideologies and cyber activities. Utilizing Python libraries, we implemented web scraping techniques to extract textual data from these Telegram channels.

**Data Scraping Process** - Python libraries such as "requests" and "BeautifulSoup" were instrumental in scraping data from Telegram channels. By sending HTTP requests to the Telegram web interface and parsing the HTML content, we retrieved textual data from targeted channels. Moreover, we employed the *telethon l*ibrary, a Python wrapper for the Telegram API, to access channel content programmatically. This facilitated the extraction of data in a structured format, ensuring consistency across datasets.

**Data Pre-processing** - To accomplish this goal, we implemented a system where each dataset includes a "class" column. Here, a value of "0" denotes data related to politics, while "1" signifies data relevant to cybersecurity. This binary classification framework is essential for aiding machine learning (ML) and neural network (NN) algorithms in understanding the content of the data during training.

Our approach involves a data preprocessing pipeline to streamline model training and evaluation. This pipeline consists of two primary stages: data partition and TF-IDF vectorization.

Initially, we segment our dataset into training and testing subsets using the `train_test_split` function from the `sklearn.model_selection` module. This step randomly divides our dataset into two parts: one for training, comprising 75% of the data, and another for testing, encompassing 25% of the data. By maintaining a consistent random state (random_state = 42), we ensure reproducibility across experiments.

Subsequently, we employ TF-IDF (Term Frequency-Inverse Document Frequency) vectorization to convert textual data into numerical representations. This is accomplished through the TfidfVectorizer from the sklearn.feature_extraction.text module. Initially, the vectorizer is fitted to the training data (x_train), which involves learning the vocabulary and computing the TF-IDF weights. Afterwards, both the training and testing textual data (x_train and x_test, respectively) are transformed into TF-IDF numerical vectors. This transformation ensures that our model can effectively interpret textual data, thereby facilitating subsequent machine learning tasks.

**Dataset Creation -** Upon scraping the data, we organized it into two distinct datasets, each featuring columns for content, date, and the name of the group. This segregation facilitated the subsequent training and evaluation of machine learning models. The division into two datasets, one focusing on cyber threats and the other on political discourse, allowed for targeted analysis by domain experts.

**Model Training and Evaluation** - For model training, we employed libraries such as TensorFlow and scikit-learn, leveraging their rich ecosystem of algorithms.

We used Feedforward Neural Network (FNN), designed for a binary classification task. The model architecture is specified using the Sequential API, consisting of three fully connected layers (Dense). The first layer has 128 neurons and an input shape corresponding to the number of features in the training dataset, using the ReLU activation function. To mitigate overfitting, a Dropout layer with a rate of 0.5 is added after the first and second dense layers. The second dense layer has 64 neurons, also using ReLU activation. The final layer consists of a single neuron with a sigmoid activation function to output a probability score for the binary classification.

The model is compiled with the Adam optimizer, binary cross-entropy loss function, and accuracy as a metric to evaluate performance. Training is conducted over 10 epochs with a batch size of 32, and the performance is validated using the test set (xv_test, y_test).

Next we used Long Short-Term Memory (LSTM) model from the Keras library, designed for a binary classification task. The model architecture is specified using the Sequential API. The first layer is an LSTM layer with 128 units, configured to process input data with a shape of (1, xv_train.shape[1]), where the input shape indicates sequences of length 1 with a number of features equal to xv_train.shape[1]. Following the LSTM layer, a Dropout layer with a rate of 0.5 is added to reduce overfitting by randomly setting half of the input units to 0 at each update during training. The next layer is a fully connected Dense layer with 64 neurons and ReLU activation function, followed by another Dropout layer with a rate of 0.5. The final layer is a Dense layer with a single neuron and a sigmoid activation function to output a probability score for the binary classification.

The model is compiled with the Adam optimizer, binary cross-entropy loss function, and accuracy as a metric to evaluate performance. Training is conducted over 10 epochs with a batch size of 32, using reshaped training data (xv_train_reshaped) and validated on reshaped test data (xv_test_reshaped).

The goal of the workflow is to successfully analyse hacktivists behaviour patterns by following the outlined steps.

## RESULTS

The results of our experiments showing the performance results scores of different classification algorithms are in Table 1: FNN (0.90), LSTM (0.89) and SVM (0.88). Despite the proximity of results, FNN stands out marginally superior. Notably, we conducted 10 tests for each algorithm and selected the highest score attained in each case.

Table 1 – Tests results

|  | **Feedforward NN** | **LSTM** | **SVM** |
|---|---|---|---|
| **Score** | 0.90 | 0.89 | 0.88 |
| **Root Mean Squared Error (RMSE)** | 0.30 | 0.32 | 0.34 |
| **Relative Absolute Error (RAE)** | 0.22 | 0.10 | 0.11 |
| **Relative Squared Error (RSE)** | 0.37 | 0.43 | 0.47 |
| **Coefficient of Determination** | 0.63 | 0.56 | 0.52 |

We proceeded with our investigations using Telegram data that hadn't been included in the two datasets used to train our algorithms. Utilizing the VADER (Valence Aware Dictionary and sEntiment Reasoner), we conducted sentiment analysis. The outcomes are encouraging, showing that sentiment analysis is yielding satisfactory results (Figure 1).

```
Message 1: Sentiment Score: -0.8464, Text: I Killmilk, represented by the organizer of the hacker group Killnet, officially take full responsibility for causing particularly serious damage to the network infrastructure of the Bulgarian corrupt government. And I declare this officially! Chief Prosecutor of the Republic of Bulgaria Ivan Geshev Fuck you!
Message 2: Sentiment Score: 0.7003, Text: South Africa was a victim of apartheid for decades. We don't want any nations experience it again. This time Israel is committing genocide against Palestinians. We are doing our best to stop Israel and Free Palestine. We tried to sue against Israel in International court, and now be keep on our try in cyber court…Long live the South Africa, long live the Palestine
```

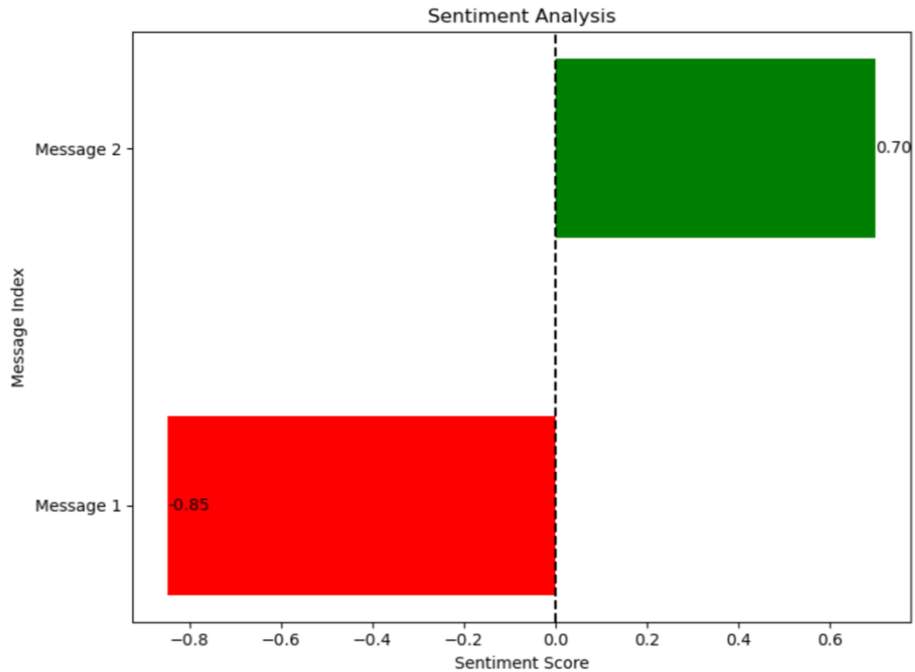

Figure 1: Results of Sentiment Analysis

Additionally, entity recognition is carried out using spaCy's named entity recognition (NER) module. For each message, entities such as persons, organizations, and geopolitical entities are extracted and stored in a structured format.

```
Entities:
Message 1:
    Person: Killmilk, Killnet
    Norp: Bulgarian
    Gpe: the Republic of, Bulgaria
Message 2:
    Gpe: South Africa, Israel, Israel, Israel, South Africa
    Date: decades
    Norp: Palestinians
```

Figure 2: Results of Entity Recognition

While we did encounter some false positive results, overall, we find the outcomes satisfactory. These results should provide analysts with enough information to discern potential vectors of attack, objectives, and relevant topics.

## DISCUSSION

The experimental findings endorse the initial inquiry regarding the application of NLP, neural networks, and machine learning for early warning detection, consistent with prior research conducted by other investigators [7, 8, 9]. Notably, our study revealed the efficacy of sentiment analysis, even in cases where negative statements contained positive words or positive statements featured numerous negative words. Despite some inaccuracies in entity recognition, particularly in identifying nouns as names, the overall test outcomes are satisfactory and will aid analysts in filtering out irrelevant data.

## CONCLUSION

The research findings highlight the promising performance of FNN, LSTM, and SVM models in analyzing data from Telegram channels for early threat detection. Neural networks, in particular, produce superior results but require more computational resources and time. The effectiveness of these models is significantly influenced by the quality of the datasets, underscoring the importance of robust dataset creation for training. Sentiment analysis and entity recognition are valuable for identifying threats and helping analysts filter out non-threatening data. The results suggest that neural networks have the potential to significantly aid cybersecurity experts in identifying threats more efficiently, leading to considerable time savings.

Interpreting these findings reveals that although neural networks perform better, the trade-off in computational resources might not always be justified, especially given the efficiency of other machine learning algorithms like SVM. This highlights the need for a balanced approach, considering both performance and resource constraints when selecting models.

Future research should focus on expanding datasets from hacktivist groups and analyzing these larger datasets with a broader range of algorithms and models beyond those used in this study. Additionally, exploring advanced techniques such as reinforcement learning or ensemble methods could further enhance threat detection capabilities. These research directions will contribute to the development of more robust and efficient threat detection systems.

## ACKNOWLEDGEMENTS

This work was supported by the National Science Program "Security and Defense", which has received funding from the Ministry of Education and Science of the Republic of Bulgaria under the grant agreement No. Д01-74/19.05.2022.